\documentclass{JHEP3}
\newcommand{\beq}{\begin{equation}}
\newcommand{\eeq}{\end{equation}}
\newcommand{\phib}{\ensuremath{\overline{\phi}}}
\newcommand{\cF}{\ensuremath{{\cal F}}}

\newcommand{\KD}{K\"{a}hler-Dirac }

\usepackage{epsfig,multicol}

\title{A geometrical approach to $N=2$ super Yang-Mills theory on the 
two dimensional lattice}

\author{Simon Catterall\\
        Department of Physics, Syracuse University, Syracuse, NY 13244, USA\\
        E-mail: \email{smc@physics.syr.edu}\\
        }

\preprint{SU-4252-803}  

\abstract{We propose a discretization of two
dimensional Euclidean Yang-Mills theories with
$N=2$ supersymmetry which preserves exactly both gauge invariance
and an element of supersymmetry. The approach
starts from the twisted form of the continuum super Yang Mills action
which we show may be written in
terms of two real \KD fields whose components
transform into each other under the twisted
supersymmetry. Once the theory is written in this
geometrical language it is straightforward to
discretize by mapping 
the component tensor fields to appropriate geometrical structures in
the lattice and by replacing the continuum exterior derivative and its adjoint
by appropriate lattice covariant difference operators. The
lattice action is local and possesses a unique vacuum state while
the use of \KD fermions ensures the model does not
exhibit spectrum doubling. 
}


\begin{document} 
\section{Introduction}
Supersymmetric field theories play a central role in modern theories
of particle physics. From a phenomenological viewpoint they are
attractive as providing a solution to the gauge hierarchy problem \cite{wein}.
From a theoretical perspective they are more tractable analytically
than their non-supersymmetric counterparts while still exhibiting
features like confinement and chiral symmetry breaking \cite{prop}.
Super Yang-Mills theories are especially interesting because of
their possible connection to string and M-theory \cite{M}. 

For these reasons a good deal of
effort has gone into attempts to formulate such theories on spacetime
lattices see, for example, \cite{old,recent} and the recent reviews
by Feo and Kaplan \cite{feo,kaplan}.  
However, until recently these efforts mostly met with only limited
success. The reasons for this are well known -- 
generic discretizations of supersymmetric field
theories break supersymmetry at the classical level leading to the
appearance of a 
plethora of {\it relevant} SUSY breaking counterterms in the effective
action. 
The couplings to all these terms must then be fine tuned as the lattice
spacing is reduced in order that the theory approach a supersymmetric
continuum limit. This problem is particularly acute in theories
with extended supersymmetry which contain scalar fields.

One might hope that this fine tuning problem might be reduced or
perhaps even eliminated by formulating the lattice models in such a way
as to preserve some element of SUSY on the lattice.
An approach following this philosophy has been described 
in papers by Kaplan et al.\cite{kap1,kap2,kap3}. 

In \cite{top} we proposed
a different scheme, useful for theories with
extended supersymmetry, based on a reformulation of the theories using
ideas drawn from topological quantum field theory. 
The key to this approach is to construct a new rotation group
from a combination of the original rotation group and part of the
R-symmetry associated with the extended SUSY. The supersymmetric field
theory is then reformulated in terms of fields which transform as
integer spin representations of this new rotation group
\cite{witten1}.
This process is given the name {\it twisting} and in flat space one
can think of it as merely an exotic change of variables in the theory.
In this process a scalar anticommuting field is always produced
associated with a nilpotent supercharge $Q$. Furthermore, as argued in
\cite{conf} the twisted superalgebra implies that
the action rewritten in terms of these twisted fields is generically
$Q$-exact. In this case it is straightforward to construct a lattice
action which is $Q$-invariant provided {\it only} that we preserve
the nilpotency of $Q$ under discretization. 
Concrete examples of this construction for theories
without gauge symmetries were given in \cite{qm,wz2,sigma}
corresponding to supersymmetric quantum mechanics, the 2D complex Wess-Zumino
model and supersymmetric sigma models. 

In \cite{joel}
the conditions allowing for a nilpotent supercharge were
analyzed in some detail within a conventional superspace
approach. In \cite{noboru} a twisted superspace 
formalism was developed and used
to construct models which preserved all the twisted supercharges at the
expense of introducing some non-commutativity at the scale of
lattice spacing.
In \cite{sug1} Sugino managed to extend the technique of latticization
via twisting
to the case of models with gauge symmetry by a non-trivial
modification of the twisted supersymmetry transformations. However, the lattice
models constructed this way have some difficulties -- they generically
suffer from a vacuum degeneracy problem\footnote{this problem was circumvented
in the case of $N=2$ SYM in two dimensions \cite{sug2}} and
the lattice actions are not rotationally invariant. Both of these problems
may be traced to the requirement that all fields {\it except} for the gauge
links transform identically under the gauge group despite
their differing spins and hence geometrical characters. We are
thus motivated to seek a more geometrical approach to discretization
of the continuum twisted theory.

In this paper we propose an alternative lattice regularization
scheme for two-dimensional $N=2$ (Euclidean) super Yang-Mills theory.
Our jumping off point is again the continuum
twisted theory. First, to show that these twisted models
are completely equivalent in flat space to the conventional formulations,
we show how to reconstruct 
the usual super Yang-Mills theory written in terms of spinor
fields from the twisted model. Next we introduce the notion of
a \KD field and recall the relationship between the \KD equation and
the usual Dirac equation. 
We show that 
the anticommuting twisted fermion fields arising in the super Yang-Mills model 
are nothing more than components of a single real \KD
field. The usual flavor index of the \KD field
is now naturally associated with an index describing the 
behavior of the field under additional R-symmetries associated
with the extended supersymmetry. This construction yields an explicit example
of the connection between twisting and the \KD fermion mechanism
emphasized in recent papers \cite{tak,kato,noboru}.
The connection to \KD fermions is important as it has been known
for some time how to discretize the latter equation
without encountering spectrum doubling \cite{rabin,betch,joos,banks}.
Indeed, we show that
the twisted lattice fermion action we propose is nothing 
more than a latticized,
gauged \KD action for fields in the adjoint representation of
the gauge group. 

Furthermore, we can show that the entire theory can be
recast as one involving a single \KD field with grassmann components
representing the twisted fermions, together
with another \KD field with commuting component fields
representing the scalars, auxiliary field and gauge field. The twisted
supersymmetry operator then induces transformations between corresponding
components of these \KD fields. 
This fully geometric representation of the continuum theory can then be
naturally discretized while preserving gauge invariance, supersymmetry and
without inducing fermion doubles. 
The discretization prescription we use was first proposed in \cite{adjoint}
and 
maps continuum fields which transform differently
under the (twisted) rotation group to different geometrical features in
the hypercubic lattice. Specifically we assign scalar fields to
sites, vector fields to links, rank 2 antisymmetric tensor fields to
plaquettes etc. These fields will then 
be taken to transform {\it differently} at finite lattice
spacing under gauge transformations. In addition we will
introduce two covariant finite difference operators which are
compatible with these differing gauge transformation properties of
the fields. They will represent the lattice analogs of
the exterior derivative and its adjoint.
Using these ingredients we will show that it is rather
straightforward to
latticize the continuum twisted theory while maintaining both
invariance under lattice gauge transformations and a single
twisted supersymmetry. The resultant action is moreover
local, has a unique vacuum state and is free of doubler modes.

\section{Two dimensional continuum $N=2$ SYM}
Our starting point will be the continuum twisted
form of the two dimensional $N=2$ SYM model
which possesses two scalar fields $\phi,\overline{\phi}$, a vector
$A_\mu$ and another commuting field $B_{12}$ corresponding to the single
independent component of a rank 2 antisymmetric tensor field in two
dimensions. The fermions of the theory appear as an anticommuting scalar
field $\eta$, a vector $\psi_\mu$ and a field $\chi_{12}$ conjugate
to $B_{12}$. All these fields are taken in the adjoint representation of
some gauge group $C=\sum_a T^aC^a$ where the $T^a$'s will be taken
to be {\it anti-hermitian} generators of the group and the component
fields $C^a$ are real. The twisted action takes the form
\beq
S=\beta Q{\rm Tr}\int
d^2x\left(
\frac{1}{4}\eta[\phi,\phib]+2\chi_{12}F_{12}+\chi_{12}B_{12}+\psi_\mu D_\mu \phib\right)
\label{gfermion}
\eeq
where the object inside the $Q$-variation we shall refer to as the
twisted gauge fermion in analogy with usual BRST terminology.
The twisted supersymmetry acts on the fields as
\begin{eqnarray}
QA_\mu&=&\psi_\mu\nonumber\\
Q\psi_\mu&=&D_\mu\phi\nonumber\\
Q\phi&=&0\nonumber\\
Q\chi_{12}&=&B_{12}\nonumber\\
QB_{12}&=&[\phi,\chi_{12}]\nonumber\\
Q\phib&=&\eta\nonumber\\
Q\eta&=&[\phi,\phib]
\end{eqnarray}
Notice that the square of twisted supersymmetry operator yields
an infinitesimal gauge transformation $Q^2=\delta_G^\phi$ with parameter $\phi$.
Carrying out the $Q$-variation and integrating out the multiplier field 
$B_{12}$ leads to the form
\begin{eqnarray}
S&=&\beta {\rm Tr}\int d^2x\left(
\frac{1}{4}[\phi,\phib]^2-\frac{1}{4}\eta [\phi,\eta]-F_{12}^2+D_\mu \phi D_\mu \phib \right. \nonumber\\
&-&\left.\chi_{12} [\phi,\chi_{12}]-
2\chi_{12}\left(D_1\psi_2-D_2\psi_1\right)-\psi_\mu D_\mu\eta+\psi_\mu [\phib,\psi_\mu]\right)
\label{twist_sym_action}
\end{eqnarray}
The coefficient of $F^2_{12}$ appears negative but this is an illusion.
With our representation of the generators ${\rm Tr}\{T_a,T_b\}=-\delta_{ab}$ and
the gauge action written in terms of component fields is positive semidefinite.  
To show that this
twisted model is nothing more than the
usual SYM theory, in which the fermions are represented by spinor fields, we construct a
Dirac spinor out the four (real) anticommuting twisted fields
\beq
\Psi=\left(\begin{array}{c}
\frac{1}{2}\eta-i\chi_{12}\\
\psi_1-i\psi_2
\end{array}\right)\eeq
It is straightforward to see that the kinetic terms in \ref{twist_sym_action} can be rewritten
in the Dirac form
\beq
\Psi^\dagger \gamma.D \Psi\eeq
where the gamma matrices are taken in the Euclidean chiral representation
\beq
\begin{array}{cc}
\gamma_1=\left(\begin{array}{cc}
0&1\\
1&0\end{array}\right)
&
\gamma_2=\left(\begin{array}{cc}
0&i\\
-i&0\end{array}\right) 
\end{array}
\eeq
In the same way the Yukawa interactions with the scalar fields can be written
\beq
\Psi^\dagger\frac{\left(1+\gamma_5\right)}{2}[\phib,\Psi]-
\Psi^\dagger\frac{\left(1-\gamma_5\right)}{2}[\phi,\Psi]\eeq
where $\gamma_5$ in this representation is
\beq
\gamma_5=\left(\begin{array}{cc}
1&0\\
0&-1\end{array}\right)\eeq
Thus the on-shell twisted action is nothing more than the usual $N=2$ SYM action in two
dimensions. Notice that to make this correspondence and
obtain a bounded Euclidean action it is necessary to think of
$\phi$ and $\phib$ as complex conjugates rather than real independent fields.
In the continuum theory this complexification is a little mysterious but
we will see that it is natural within a lattice framework.
Finally with $\phi$ and $\phib$ complex conjugates
it is easy to show that $\gamma_1 M^*\gamma_1=M$ where $M$ is
the fermion operator which implies that the fermion determinant is (generically)
positive definite.

\section{Interpretation in terms K\"{a}hler-Dirac fields}
The fact that the fermions of the twisted model are represented
by (antisymmetric) tensor
fields is reminiscent of the component fields entering
into the \KD equation. In this section we verify this
connection by
showing how to write the original twisted SYM theory 
entirely in terms of such \KD fields. We start by recalling the properties of
the \KD equation and its connection to spinor fields and
the usual Dirac equation \cite{rabin,joos,banks}

In
$D$ dimensions we introduce a \KD field $\omega$ whose components are
antisymmetric tensor fields or p-forms where $p=0\ldots D$. Thus
\beq
\omega=\left(f,f_\mu,f_{\mu\nu},\ldots\right)\eeq
We can define the action of the exterior derivative $d$ on such a field by the
action on its components
\beq
d\omega = \left(0,\partial_\mu f, \partial_\mu f_\nu - \partial_\nu
f_\mu,\ldots\right)\eeq
A natural dot product between two such \KD fields $A$ and $B$ is given by
\beq<A|B>=\int d^Dx\sqrt{g}\sum_p A^{\mu_1\ldots\mu_p}B_{\mu_1\ldots\mu_p}\eeq
The adjoint of the exterior derivative $d^\dagger$ can then be defined in terms of
the component fields as 
\beq-d^\dagger\omega=\left(f^\nu,f^\nu_\mu,\ldots,0\right)_{;\nu}\eeq
If we form the matrix (from now on we will assume flat Euclidean
space)
\beq
\Psi^{(\omega)}_{\alpha\beta}(x)=\sum_{p=0}^D
\left(\gamma^{\mu_1}\ldots\gamma^{\mu_p}\right)_{\alpha\beta}
f_{\mu_1\ldots\mu_p}\eeq
it is straightforward to show that that the following equation 
\beq
\gamma^\mu_{\alpha\alpha^\prime}\partial_\mu
 \Psi^{(\omega)}_{\alpha^\prime\beta}=0\label{matrixeq}\eeq
is equivalent to the \KD equation
\beq
(d-d^\dagger)\omega=0\eeq
Furthermore we can interpret eqn.~\ref{matrixeq} as the usual Dirac
operator acting on a multiplet of $2^{\frac{D}{2}}$ identical
flavors of Dirac fermions labeled by the index $\beta$.  
This statement is unaffected if gauge interactions are introduced
and the derivative operator $d$ replaced by an appropriate gauge covariant
exterior derivative $D$.
Furthermore, the \KD equation can be 
derived from an action which can be written in two equivalent ways
\begin{eqnarray*}
S&=&\frac{1}{2}{\rm Tr}\Psi^\dagger \gamma_\mu \partial_\mu\Psi\nonumber\\
 &=&\left<\omega^\dagger| \left(d-d^\dagger\right)\omega\right>
\end{eqnarray*}
where the (anticommuting) matrix valued fields $\Psi$ and $\Psi^\dagger$ (and
correspondingly the \KD fields $\omega$, $\omega^\dagger$) are to be
treated as independent fields.
To make contact with the twisted SYM model
discussed earlier let us examine in detail the
case of two dimensions. Choosing
\beq
\omega=\left(\frac{\eta}{2},\psi_\mu,\chi_{12}\right)\eeq
we find the the continuum \KD action when expanded on the 
component fields yields
\beq
S=\int d^2x \left(
\frac{1}{2}\eta^\dagger D_\mu\psi_\mu+\frac{1}{2}\psi^\dagger_\mu D_\mu \eta+
\psi_2^\dagger D_1\chi_{12}-\psi_1^\dagger D_2\chi_{12}+
\chi^\dagger_{12}\left(D_1\psi_2-D_2\psi_1\right)\right)
\eeq
If we now impose the condition that the component fields are purely
anti-hermitian (or hermitian) we obtain the continuum twisted $N=2$
action examined earlier (up to an unimportant factor of
minus two). Notice that in this case the total
number of real fields needed to write down the \KD model (four) exactly
matches the number of real supercharges of the $N=2$ theory in
two dimensions.
We thus see that the fermionic sector of the $N=2$ model in
two dimensions is naturally given in terms of a single real \KD fermion.
It is clear that
the bosonic sector of the model contains a similar set of
four real fields $\phib$, $A_\mu$ and $B_{12}$ together with the
gauge degree of freedom $\phi$.
Therefore let us introduce another \KD field
$\Phi=(\phib-\phi,A_\mu,B_{12})$
with commuting components.
Consider the following expression
\beq
\left<\Psi|\left(d_A \Phi + Q\Psi\right)\right>
\label{DKfermion}\eeq
where $d_A$ denotes the covariant form of $d$.
Writing out components yields
\beq
\int d^2x\left(\psi_\mu D_\mu\phib+\chi_{12}2F_{12}+
\frac{1}{4}\eta[\phi,\phib]+
\chi_{12}B_{12}\right)\eeq
This is nothing more than the gauge fermion used in the continuum
twisted SYM model.
Actually we should be a little careful here -- to derive the correct
fermionic \KD action we should really introduce two independent \KD fields
$\Psi$ and $\Psi^\dagger$. This necessitates using a gauge fermion
of the form
\beq
\frac{1}{2}\left<\Psi^\dagger|\left(d_A\Phi+ Q\Psi\right)\right>+
\frac{1}{2}\left<\left(d_A\Phi+Q\Psi\right)^\dagger|\Psi\right>
\label{kdgaugefermion}
\eeq
There are two ways to reduce this theory to the usual Yang-Mills
model.  We have already seen
one simple method -- assume we can
impose a reality condition on the fields after
$Q$-variation. 
This is what we shall do later in the lattice theory.
However, in the continuum there is another way to proceed
by requiring that the
fields $\Psi$ and $\Psi^\dagger$ appearing in the gauge fermion
eqn.~\ref{kdgaugefermion} be replaced by {\it self-dual} fields
$\Psi_+$ and $\Psi^\dagger_+$ where 
\beq
\Psi_+=P_+\Psi\eeq
and the projection operator is given by $P_+=\frac{I+*}{2}$. The $*$ symbol
denotes a duality operation (related
to the Hodge dual) taking $p$ forms into $(D-p)$ forms.
For a p-form $A$
the associated dual $(D-p)$-form has components
\beq
A_{\nu_1\ldots\nu_{D-p}}=i(-1)^{\frac{1}{2}(D-p)(D-p+1)}\epsilon_{\mu_1\ldots\mu_p|\nu_1\ldots\nu_{D-p}|}
A_{\mu_1\ldots\mu_p}\eeq
where the notation $|\mu_1\ldots\mu_p|$ means only terms with 
$\mu_1<\mu_2<\cdots\mu_p$ are included in the sum. The tensor $\epsilon$ is
the completely antisymmetric symbol in $D$ dimensions. In the matrix
language the projection corresponds to right multiplication by
the matrix
$\frac{(1+\gamma_5)}{2}$.
The resulting \KD matrices contain a single non-zero column corresponding
to a single Dirac spinor and the reduction is complete.

The above analysis shows that the usual twisted SYM theory
can be elegantly rewritten in the language of \KD fields. This
rewriting of the theory in terms of differential forms has two
primary advantages -- it allows us to formulate the theory on
a curved space and also gives a natural starting point for
discretization. Indeed it has been
shown \cite{rabin,joos,betch} that any non-gauge theory formulated
in such geometrical terms may be discretized on
a hypercubic lattice {\it without inducing
fermion doubling} by replacing the exterior derivative
$d$ by a forward difference operator $D^+_\mu$ and its adjoint $d^\dagger$
by a backward difference $-D^-_\mu$. Furthermore, in \cite{adjoint}, it was
shown how to construct covariant versions of these difference operators
for fields taking their values in the adjoint representation of
a gauge group. It is hence natural to try to use \KD fields to
formulate lattice supersymmetric actions. Attempts were made to
construct such theories in a Hamiltonian formalism in \cite{n=1n=2,n=4}.
A similar approach was used in \cite{n=2sym} to construct
a SYM model in Euclidean space using only some of the
component \KD fields. 
However it is only in the context of twisted supersymmetry that
the full power of the \KD approach can be realized.

\section{General prescription for discretization}
We list here for reference the essential
ingredients in our discretization prescription. Notice they do not depend
on dimension.
\begin{itemize}
\item A continuum p-form field $f_{\mu_1\ldots \mu_p}(x)$ will be mapped
to a corresponding lattice p-form field associated with
the $p$-dimensional hypercube at lattice site $x$ 
spanned by the (positively directed) unit vectors $\{\mu_1\ldots\mu_p\}$. 
\item Such a lattice
field will transform under gauge transformations\footnote{we use $G^{-1}$
rather than $G^\dagger$ to allow us to consider complexified
gauge transformations later -- we thank Joel Giedt for this suggestion} 
in the following way
\beq
f_{\mu_1\ldots\mu_p}(x)\to G(x)f_{\mu_1\ldots \mu_p}(x)G^{-1}(x+e_{\mu_1\ldots\mu_p})\eeq
where the vector $e_{\mu_1\ldots\mu_p}=\sum_{j=1}^p\mu_j$. 
\item To construct gauge invariant quantities we will need to introduce both
$f_{\mu_1\ldots\mu_p}$ 
and its hermitian conjugate $f^\dagger_{\mu_1\ldots\mu_p}(x)$. The
latter transforms as
\beq
f^\dagger_{\mu_1\ldots\mu_p}(x)\to G(x+e_{\mu_1\ldots\mu_p})
f_{\mu_1\ldots \mu_p}(x)G^{-1}(x)\eeq
This differing transformation law for the field and its adjoint requires
that the component fields $f_{\mu_1\ldots\mu_p}^a$ be treated as
complex. 
This complexification of the degrees of freedom can be extended to scalar
fields {\it provided} they are required to be (anti)self-conjugate $f^\dagger=-f$.
Notice that such a definition departs from the usual notion of
hermitian conjugation but is natural if we want to consider a theory
with a complexified gauge invariance.
\item For a continuum gauge field we introduce lattice link 
fields $U_\mu(x)=e^{A_\mu(x)}$ 
and its conjugate
$U^\dagger_\mu=e^{A^\dagger_\mu(x)}$. 
\item A covariant forward difference operator can be defined which acts on a field
$f_{\mu_1\ldots\mu_p}(x)$
as follows
\beq
D^+_\mu f_{\mu_1\ldots\mu_p}(x)=
U_\mu(x)f_{\mu_1\ldots\mu_p}(x+\mu)-
f_{\mu_1\ldots\mu_p}(x)U_\mu(x+e_{\mu_1\ldots\mu_p})\eeq
This operator acts like a lattice exterior derivative with
respect to gauge transformations in mapping a $p$-form lattice field
to a $(p+1)$-form lattice field.
\item Similarly we can define an adjoint operator $D^-_\mu$ whose action on some field
$f_{\mu_1\ldots\mu_p}$ is given by
\beq
D^-_\mu f_{\mu_1\ldots\mu_p}(x)=
f_{\mu_1\ldots\mu_p}(x)U^\dagger_\mu(x+e_{\mu_1\ldots\mu_p}-\mu)-
U^\dagger_\mu(x-\mu)f_{\mu_1\ldots\mu_p}(x-\mu)\eeq
\item To discretize the continuum theory formulated in geometrical language simply
map all p-form fields to lattice fields as
described above and replace all instances of $d$ by $D^+$ and $d^\dagger$ by $D^-$.
\item In the final path integral we choose a contour along on which the imaginary
part of the gauge field is zero and such that the action is real, positive definite.
\end{itemize}

\section{Two dimensional lattice $N=2$ SYM}
We start from an
expression for the lattice gauge fermion which is
identical to the continuum one eqn.~\ref{kdgaugefermion}
\beq
\frac{1}{2}\left<\Psi^\dagger|\left(D\Phi+ Q\Psi\right)\right>+\frac{1}{2}
\left<\left(D\Phi+Q\Psi\right)^\dagger|\Psi\right>
\label{latticegaugefermion}
\eeq
where, following 
our discretization prescription, 
a lattice \KD field is composed of (complex-valued) p-form fields defined
on p-dimensional hypercubes in the lattice 
\begin{eqnarray*}
\Phi&=&\left(\phib-\phi,U_\mu,B_{12}\right)\nonumber\\
\Psi&=&\left(\frac{1}{2}\eta,\psi_\mu,\chi_{12}\right)
\end{eqnarray*}
together with the conjugate fields $\Phi^\dagger$ and $\Psi^\dagger$.
The fields possess the gauge
transformation properties listed in the previous section. Thus a
site, link and plaquette field transform under a gauge transformation
$G(x)=e^{\phi(x)}$ 
as 
\begin{eqnarray*}
f(x)&\to& G(x)f(x)G^{-1}(x)\nonumber\\
f_\mu(x)&\to& G(x)f_\mu(x)G^{-1}(x+\mu)\nonumber\\
f_{\mu\nu}(x)&\to& G(x)f_{\mu\nu}(x)G^{-1}(x+\mu+\nu)
\end{eqnarray*} while their conjugates transform in the
complementary way
\begin{eqnarray*}
f^\dagger(x)&\to& G(x)f(x)G^{-1}(x)\nonumber\\
f^\dagger_\mu(x)&\to& G(x+\mu)f_\mu(x)G^{-1}(x)\nonumber\\
f^\dagger_{\mu\nu}(x)&\to& G(x+\mu+\nu)f_{\mu\nu}(x)G^{-1}(x)
\end{eqnarray*}
While we are free to regard the site fields $\phi$, $\phib$ and
$\eta$ as complex-valued the above
transformations require them to satisfy an (anti)self-conjugacy condition eg.
$\phib^\dagger=-\phib$. Notice that this requirement is consistent with
the promotion of the original gauge invariance to invariance under
the complexified group -- the lattice gauge
transformation $G^\dagger=G^{-1}$ if $\phi^\dagger=-\phi$.
It is clear that these transformations reduce to
the usual ones for connections and fields in the adjoint of the
gauge group in the naive continuum limit.
As usual we
regard the gauge link as the exponential of
some matrix $U_\mu(x)=e^{A_\mu(x)}$ where
$A_\mu(x)$ and indeed all other lattice fields may be expanded on
a basis of (anti-hermitian) traceless generators of the gauge
group $f(x)=\sum_a f^a(x)T^a$.  
The doubling of degrees of freedom in the lattice theory is, at
first sight, a little puzzling -- clearly in the fermionic sector it
is nothing more than the usual statement
that the spinors $\psi$ and $\overline{\psi}$ are to
considered as independent in Euclidean
space. However, in a model with
twisted supersymmetry this necessarily seems to imply a
corresponding doubling of bosonic states.
Another way to understand this doubling 
of $p$-form fields is to recognize that it can be taken to represent
the two possible orientations of the underlying $p$-dimensional hypercube.
The complexification of
the vector potential $A_\mu^a(x)$ has the benefit of allowing
the fields $U(x)$ and $U^\dagger(x)$ to vary independently
under the twisted supersymmetry. 
In the end we will require 
the final path integral be taken along a contour where
$U^\dagger U=I$ and the imaginary parts of
the gauge field and the fermion fields vanish. This reality
condition will allow contact to be made with the usual
twisted continuum theory.

Returning now to the expression for the lattice gauge fermion in \KD
language we
define the lattice covariant exterior derivative $D$ acting
on \KD fields 
in terms of the action of $D^+_\mu$ on the component fields 
\beq
D\Phi=\left(0,D^+_\mu\left(\phib-\phi\right),2\cF_{12}\right)\nonumber\\
\eeq
where, from our discretization rules the action of 
the covariant finite difference operator $D^+_\mu$ 
on a site field $f(x)$ and a link field $f_\mu(x)$ are given explicitly
by 
\begin{eqnarray*}
D^+_\mu f(x) &=& U_\mu(x)f(x+\mu)-f(x)U_\mu(x)\nonumber\\
D^+_\mu f_\nu(x)&=& U_\mu(x)f_\nu(x+\mu)-f_\nu(x)U_\mu(x+\nu)
\end{eqnarray*}
The plaquette field $\cF_{12}$ is thus given by
\beq
\cF_{12}(x)=D^+_1 U_2(x)=U_1(x)U_2(x+1)-U_2(x)U_1(x+2)\eeq
Notice that it is automatically antisymmetric in its indices and reduces to
the usual Yang-Mills field strength in the continuum limit $a\to 0$.
Using these rules we can
now write down the form of the lattice gauge fermion
in terms of component fields.
The result, which is explicitly gauge invariant
and reduces to the continuum expression eqn.~\ref{gfermion}
in the naive continuum limit, is
\begin{eqnarray}
S_L&=&\beta Q{\rm Tr}\sum_{x}\left(\frac{1}{4}\eta^\dagger(x)[\phi(x),\phib(x)]+
\chi^\dagger_{12}(x)\cF_{12}(x)+\chi_{12}(x)\cF_{12}(x)^\dagger\right.\nonumber\\
&+&\left.\frac{1}{2}\chi^\dagger_{12}(x)B_{12}(x)+\frac{1}{2}\chi_{12}(x)B^\dagger_{12}(x)+
\frac{1}{2}\psi^\dagger_\mu(x)D^+_\mu\phib(x)+\frac{1}{2}\psi_\mu(x)(D^+_\mu\phib(x))^\dagger\right)
\end{eqnarray}
This expression will also be $Q$-invariant if we can generalize the
continuum twisted supersymmetry transformations in such a way
that we preserve the property $Q^2=\delta_G^\phi$. The following transformations
do the job
\begin{eqnarray}
QU_\mu&=&\psi_\mu\nonumber\\
Q\psi_\mu&=&D^+_\mu\phi\nonumber\\
Q\phi&=&0\nonumber\\
Q\chi_{12}&=&B_{12}\nonumber\\
QB_{12}&=&[\phi,\chi_{12}]^{(12)}\nonumber\\
Q\phib&=&\eta\nonumber\\
Q\eta&=&[\phi,\phib]
\end{eqnarray}
where the superscript notation indicates a {\it shifted} commutator
\beq
[\phi,\chi_{\mu\nu}]^{(\mu\nu)}=\phi(x)\chi_{\mu\nu}(x)-
\chi_{\mu\nu}(x)\phi(x+\mu+\nu)\eeq
These arise naturally when we consider the infinitesimal 
form of the gauge transformation
property of the plaquette field. Notice that gauge invariance also
dictates that we must use the covariant forward difference operator $D^+_\mu$ on the
right-hand side
of the $U_\mu$ variation. The $Q$-transformations of the  
conjugate fields
are similar
\begin{eqnarray}
QU^\dagger_\mu&=&\psi^\dagger_\mu\nonumber\\
Q\psi^\dagger_\mu&=&(D^+_\mu\phi)^\dagger\nonumber\\
Q\chi^\dagger_{12}&=&B^\dagger_{12}\nonumber\\
QB^\dagger_{12}&=&\left([\phi,\chi_{12}]^{(12)}\right)^\dagger\nonumber\\
\end{eqnarray}
Carrying out the $Q$-variation leads to the following expression
for the lattice action
\begin{eqnarray}
S_L&=&\beta{\rm Tr}\sum_x \left(
\frac{1}{4}[\phi(x),\phib(x)]^2-\frac{1}{4}\eta^\dagger(x)[\phi(x),\eta(x)]-
\chi^\dagger_{12}(x) [\phi(x),\chi_{12}(x)]^{(12)}+B^\dagger_{12}(x) B_{12}(x)\right.\nonumber\\
&+&B^\dagger_{12}(x)\cF_{12}(x)+B_{12}(x)\cF_{12}(x)^\dagger+
\frac{1}{2}(D^+_\mu \phi(x))^\dagger D^+_\mu \phib(x)+
\frac{1}{2}D^+_\mu \phi(x)(D^+_\mu\phib(x))^\dagger\nonumber\\
&-&\chi^\dagger_{12}(x)D^+_1\psi_2(x)+\chi^\dagger_{12}(x)D^+_2\psi_1(x))-
   \psi^\dagger_{2}(x)D^-_1\chi_{12}(x)+\psi^\dagger_1(x)D^-_2\chi_{12}(x)\nonumber\\
&-&\left.\frac{1}{2}\psi^\dagger_\mu(x)D^+_\mu\eta(x)-
         \frac{1}{2}\eta^\dagger(x) D^-_\mu \psi_\mu(x)+
\psi^\dagger_\mu(x) [\phib(x),\psi_\mu(x)]^{(\mu)}\right)
\label{twist_sym_latt_action}
\end{eqnarray}
Notice in this expression the appearance of the covariant backward 
difference operator $D^-_\mu$ whose action on a plaquette field $f_{\mu\nu}$ is
given explicitly by
\beq
D^-_\mu f_{\mu\nu}(x)=f_{\mu\nu}(x)U^\dagger_\mu(x+\nu)-U^\dagger_\mu(x-\mu)f_{\mu\nu}(x-\mu)
\eeq
the resulting object transforming as a link field under gauge
transformations. 
Similarly, following our discretization
prescription, the lattice covariant difference operator
$D^-_\mu$ acting on a link field yields
\beq
D^-_\mu f_\mu(x)=f_\mu(x)U^\dagger_\mu(x)-U^\dagger_\mu(x-\mu)f_\mu(x-\mu)
\eeq
and transforms as a site field under gauge transformations.
Notice that the $Q$-variation of $F_{12}(x)$ yields a derivative term on 
the anticommuting fields completely analogous to that
seen in the continuum. In addition 
the (link) shifted commutator term $\psi^\dagger_\mu[\phib,\psi_\mu]$
appears naturally from the variation of the last terms in the lattice
gauge fermion.
Finally we must integrate out the multiplier fields $B_{12}$ and $B^\dagger_{12}$
resulting in the term
\beq
\beta {\rm Tr}\sum_x \cF_{12}(x)^\dagger\cF_{12}(x)\eeq
This can be written
\beq
\beta{\rm Tr}\sum_x \left(2I-U_P-U^\dagger_P\right)+
\beta{\rm Tr}\sum_x \left(M_{12}+M_{21}-2I\right)
\eeq
where \beq
U_P={\rm Tr}\left(U_1(x)U_2(x+1)U^\dagger_1(x+2)U^\dagger_2(x)\right)\eeq
resembles the
usual Wilson plaquette operator
and 
\beq M_{12}(x)=U_1(x)U^\dagger_1(x)U_2(x+1)U^\dagger_2(x+1)\eeq
Notice that the second term vanishes when the gauge
field is restricted to be unitary which is equivalent to 
requiring
${\rm Im}A_\mu(x)=0$. In this case the action is nothing more than
the usual Wilson gauge
action and {\it does not} suffer from the vacuum degeneracy problem
inherent in the models constructed in \cite{sug1}.

Having constructed the lattice action we must now discuss the path
integral we will use to define the quantum theory. Initially
this path integral will include integrations over both fields
and their conjugates. To make contact with the continuum theory
we would like to integrate along a contour on which the
imaginary parts of all fields bar the scalars vanish (and the
scalars are taken to be hermitian conjugates of each other).
It is clear that the Yang-Mills action, the scalar action and
the determinant resulting from integration over the
twisted fermions are still gauge invariant when so restricted.
Furthermore it is
clear that the resulting action is real and positive definite
(at least for small lattice spacing when the lattice action approaches
the continuum action) for such
a choice of contour. The only remaining question relates to
the $Q$-symmetry - specifically do the
the twisted supersymmetric
Ward identities still hold when the lattice theory is restricted in this
way ? To see that this is the case remember that the action is
$Q$-exact and hence any supersymmetric
Ward identity can be computed {\it exactly}
in the limit $\beta\to\infty$. But in such a limit
I can expand the gauge links to leading order in $A_\mu$ 
and recover the (complexified) continuum action
and $Q$-transformations. Furthermore, it is known that the continuum theory
can be consistently restricted to the contour we have
described \cite{witten1} and so we infer that the lattice Ward
identities should also be satisfied on this contour.

Returing to the lattice action given in
eqn.~\ref{twist_sym_latt_action} we may rewrite the
fermionic pieces in the form 
\beq
\overline{\Psi}M\Psi\eeq
where $M$ the matrix operator can be written in block form
\beq
M=\left(\begin{array}{cc}
-[\phi,]^{(p)}&K\\
-K^\dagger&[\phib,]^{(p)}
\end{array}\right)\eeq
and the kinetic operator $K$ is given by
\beq
K=\left(\begin{array}{cc}
D^+_2&-D^+_1\\
-D^-_1&-D^-_2
\end{array}\right)
\label{latticeop}\eeq
with the spinors defined as 
\beq
\begin{array}{cc}
\overline{\Psi}=\left(
\chi^\dagger_{12},
\eta^\dagger,
\psi_1^\dagger,
\psi_2^\dagger\right)
&
\Psi=\left(\begin{array}{c}
\chi_{12}\\
\eta\\
\psi_1\\
\psi_2\end{array}\right) 
\end{array}
\eeq
Notice that the form of the kinetic operator $K$ ensures that the
lattice theory does not exhibit spectrum doubling. The absence of doubles
is not an accident but, as advertised, is a consequence of 
discretizing a purely geometrical action written in terms of
\KD fields.
Finally it is not possible at non-zero
lattice spacing to cast the theory in terms of a single Dirac
spinor. One easy way to see this is to recall that
the components of the continuum Dirac spinor contain
objects like $(\frac{1}{2}\eta+i\chi_{12})$. In the
language of \KD fields such quantities arise after the self-dual
projection. They are problematic on the lattice since they
do not 
transform simply at finite lattice spacing under gauge
transformations. Thus a reduction to a single Dirac spinor is
not possible in the lattice theory.
Instead after integrating out the anticommuting degrees of freedom
along the contour ${\rm Im}\Psi^a=0$ 
we will be left with a factor
\beq{\rm Pf}(M)\eeq
In the continuum limit we know that this Pfaffian is equivalent
to a real, positive definite determinant. Thus from the
point of view of simulations it should be possible to
replace the Pfaffian by the expression
\beq
{\rm Pf}(m)={\rm det}^{\frac{1}{2}}M\eeq
without encountering a sign problem for small enough lattice spacing.

\section{Conclusions}
In this paper we have derived a lattice action for $N=2$ super Yang-Mills
theory in two dimensions. We first show that the continuum form of
the action can be written succinctly in the language of differential forms
and \KD fields. This manifestly geometric starting point allows us
to discretize the theory without inducing spectrum doubling and
maintaining both gauge invariance and a single twisted supersymmetry.
The lattice theory naturally contains complex fields -- to access
the correct continuum limit requires that an
appropriate contour be chosen when evaluating the path integral.
We argue that both gauge invariance and the twisted supersymmetry
can be maintained if this contour is chosen such that the
imaginary parts of all component field bar the scalars are taken
to vanish. The scalars $\phi^a(x)$ and $\phib^a(x)$ can be taken
to be (minus) the complex conjugate of each other.
The resulting fermion operator can be shown to be positive definite
at least for small enough lattice spacing.

There are several directions for further work. 
The most obvious is the need for numerical simulations to
check some of the conclusions of this work, perhaps most
importantly, the claim that the twisted Ward identities
are maintained along the contour required to
define the path integral. It is also possible to
generalize these ideas to four dimensions. The $N=4$ theory
contains $16$ real supercharges which, with an appropriate twist,
can be represented using a four dimensional (real) \KD field.
Furthermore, it is possible to embed the bosonic degrees of
freedom of this theory in another real
\KD field with commuting components just as
for the two dimensional theory. Derivation of the appropriate
gauge fermion and corresponding
twisted supersymmetry will be presented elsewhere
\cite{catt}. Such a formulation should allow for discretization
using the prescription described here.
Secondly, the geometric nature of these theories should allow them to
be formulated on arbitrary simplicial lattices
\cite{geo,catt2}. This would allow
study of twisted super Yang-Mills theories on curved spaces. Summing
over such simplicial lattices may provide a connection to 
(lattice regulated) supergravity theories.

\acknowledgments
This work
was supported in part by DOE grant DE-FG02-85ER40237. The author
would like to acknowledge useful discussions with Joel Giedt
and Noboru Kawamoto.


\begin{thebibliography}{99}
\bibitem{wein} S. Weinberg, The Quantum Theory of Fields III (Cambridge University Press, 2000) 
\bibitem{prop} N. Seiberg and E. Witten, Nucl. Phys. B431 (1994) 484.
\bibitem{M} J.M. Maldecena, Adv. Theor. Math. Phys. 2 (1998) 231 [Int. J. Theor.
Phys 38 (1999) 1113].
\bibitem{old} M Golterman and D. Petcher, Nucl. Phys. B319 (1989) 307.\\
S. Elitzur and A. Schwimmer, Nucl. Phys. B226 (1983) 109.\\
N. Sakai and M. Sakamoto, Nucl. Phys. B229 (1983) 173.\\
J. Bartels and J. Bronzan, Phys. Rev. D28 (1983) 818\\
T. Banks and P. Windey, Nucl. Phys. B198 (1983) 68
\bibitem{recent} Y. Kikukawa and Y. Nakayama, Phys. Rev. D66 (2002) 094508.\\
Kazuo Fujikawa, Phys. Rev. D66 (2002) 074510.\\
J. Nishimura, S. Rey and F. Sugino, JHEP 0302 (2003) 032\\
J. Giedt, E. Poppitz and M. Rozali, JHEP 0303 (2003) 035\\
J. Giedt, Nucl. Phys.B 668 (2003) 138\\
J. Giedt, The fermion determinant in (4,4) 2d
lattice super Yang Mills, hep-lat/0307024 \\
J. Nishimura, Phys. Lett. B406 (1997) 215\\
S. Catterall and S. Karamov, Phys. Rev. D 68 (2003) 014503\\
W. Bietenholtz, Mod. Phys. Lett. A14 (1999) 51.
\bibitem{feo} A. Feo, Supersymmetry on the Lattice, hep-lat/0210015\\
\bibitem{kaplan} D. B. Kaplan, Recent Developments in Lattice Supersymmetry,
hep-lat/0309099
\bibitem{kap1}
D.B. Kaplan, E. Katz and M. Unsal, JHEP 0305 (2003) 037.
\bibitem{kap2} A.G. Cohen, D.B. Kaplan, E. Katz, M. Unsal, 
JHEP 0308 (2003) 024   
\bibitem{kap3} A.G. Cohen, D.B. Kaplan, E. Katz, M. Unsal, hep-lat/0307012.
\bibitem{top} S. Catterall, JHEP 0305 (2003) 038. 
\bibitem{witten1} E. Witten, Comm. Math. Phys. 117 (1988) 353.
\bibitem{conf} S. Catterall hep-lat/0409015.
\bibitem{qm}
S. Catterall and E. Gregory, Phys. Lett. B487 (2000) 349.
\bibitem{wz2}
S. Catterall and S. Karamov, Phys. Rev. D65 (2002) 094501.
\bibitem{sigma} S. Catterall and S. Ghadab, JHEP 0405 (2004) 044.
\bibitem{joel} J. Giedt, Erich Poppitz, hep-lat/0407135.
\bibitem{noboru} A. D'Adda, I. Kamamori, N. Kawamoto, K. Nagata,
hep-lat/0406029.
\bibitem{sug1} F. Sugino, JHEP 0401 (2004) 015.
\bibitem{sug2} F. Sugino, JHEP 0403 (2004) 067.
\bibitem{tak} N. Kawamoto , T. Tsukioka, Phys. Rev. D61 (105009) 2000.
\bibitem{kato} J. Kato, N. Kawamoto and Y. Uchida, Int. J. Mod. Phys. A19 (2004)
2149.
\bibitem{rabin} J. Rabin, Nucl. Phys. B201 (1982) 315.
\bibitem{betch} P. Becher, Phys. Lett. B104 (1981) 221.
\bibitem{joos} P. Becher and H. Joos, Z. Phys. C 15 (1982) 343.
\bibitem{banks} T. Banks, Y. Dothan and D. Horn, Phys. Lett. B117 (1982) 413.
\bibitem{adjoint} H. Aratyn, M. Goto and A.H. Zimerman, Nuovo Cimento A84,
(1984) 255
\bibitem{n=1n=2} H. Aratyn, M. Goto and A.H. Zimerman, Nuovo Cimento A88 
\bibitem{n=4} H. Aratyn and A.H. Zimerman, J. Phys. A 18 (1985) L487.
\bibitem{n=2sym} H. Iwamoto, H. Aratyn and A. Zimerman, Z. Phys. C 31. (1986)
99.
(1985) 225.
\bibitem{catt} S. Catterall, preparation.
\bibitem{geo} H. Aratyn and A.H. Zimerman, Phys. Rev. D33 (1986) 2999.
\bibitem{catt2} S. Catterall, in preparation.
\end{thebibliography}
\end{document}